\documentclass[a4paper]{jpconf}
\usepackage{graphicx}
\begin{document}
\title{Giant monopoles as a dark matter candidate}

\author{Jarah Markar Evslin}

\address{TPCSF, Institute of High Energy Physics, CAS, YuQuanLu 19B, 100049 Beijing, China}

\ead{jarah@ihep.ac.cn}

\begin{abstract}
The most prominent successes of weakly interacting massive particle (WIMP) models of dark matter are  at comoving scales of megaparsecs.  At kiloparsec scales they face challenges in explaining the density profiles, abundances and phase space distributions of satellite galaxies in our local group.  An alternate dark matter candidate, the giant 't Hooft-Polyakov monopole, is proposed which may share the large distance successes of WIMPs while evading their short distance problems.  These are classical field theory solutions of a dark sector including a nonabelian gauge field, an adjoint scalar field and fundamental fermions.  In such models each halo consists of a single monopole characterized by a conserved integer charge.

%All articles {\it must} contain an abstract. This document describes the  preparation of a conference paper to be published in \jpcs\ using \LaTeXe\ and the \cls\ class file. The abstract text should be formatted using 10 point font and indented 25 mm from the left margin. Leave 10 mm space after the abstract before you begin the main text of your article. The text of your article should start on the same page as the abstract. The abstract follows the addresses and should give readers concise information about the content of the article and indicate the main results obtained and conclusions drawn. As the abstract is not part of the text it should be complete in itself; no table numbers, figure numbers, references or displayed mathematical expressions should be included. It should be suitable for direct inclusion in abstracting services and should not normally exceed 200 words. The abstract should generally be restricted to a single paragraph. Since contemporary information-retrieval systems rely heavily on the content of titles and abstracts to identify relevant articles in literature searches, great care should be taken in constructing both.
\end{abstract}

\section{Successes of WIMP models}

The most successful dark matter model is the cold dark matter (CDM) WIMP.  These are nonrelativistic particles which interact gravitationally at long distances and via weak interactions at short distances.  While experimental searches for their weak interactions have yielded no convincing signals \cite{lux},  the consequences of their gravitational interactions are regularly verified.  

Perhaps the most famous of these confirmations comes from the bullet cluster:  Two clusters of galaxies have passed through each other, leaving most of their luminous matter in the form of a hot plasma at the collision point.  Via gravitational lensing it was shown in Ref.~\cite{bullet} that most of the matter continued past the collision point, separating from the plasma.  The fact that the dark matter separated from the visible matter means that it has its own inertia.  This observation eliminates the most elegant modified gravity explanations of dark matter.  In addition, the fact that the dark matter halos of the clusters passed through each other yields a strong upper bound on the dark matter self-interaction cross section \cite{bulletlim}.  It has been claimed \cite{nobec} that this observation rules out a large class of models in which, like the model presented below, dark matter consists of classical field theory solutions such as the scalar condensate model of Ref.~\cite{bec}.

Once structure formation begins, CDM WIMPs are nonrelativistic and only interact gravitationally.  Based on numerical simulations it was claimed in Ref.~\cite{nfw} that, as a result of these two facts, the shape of a pure dark matter halo assumes a universal form which is well-approximated by the NFW profile.  Observations of galaxy clusters confirm that the NFW profile indeed provides an excellent approximation to the dark matter density profile \cite{cluster}.

%The power spectrum of the cosmic microwave background radiation (CMB) provides a very different and complimentary probe of the gravitational interactions of WIMPs.  
The CMB power spectrum observed up to multipoles of about $l=2000$ \cite{planck} is just what would arise from an initial power law spectrum of adiabatic fluctuations in a mixture of baryonic matter, photons and a nonrelativistic fluid in a universe with a cosmological constant and massless neutrinos.  Around $l=2000$, combining CMB data with other data, the interpretation becomes more controversial \cite{spt} and above $l=3000$ the primordial fluctuations have been erased by Silk damping \cite{silk}.  Therefore CMB observations demonstrate that up to redshifts of about $10^4$ and at angular scales down to about $l=2000$ dark matter behaves as a perfect fluid with an equation of state of about $w=0$.

The above evidence for dark matter is subject to various assumptions and unknown systematics.  For example, since the power spectrum of primordial perturbations is entirely unknown there is no way to estimate the error caused by its deviation from a power law.  While the bullet cluster seems to be a convincing demonstration of the existence of particulate dark matter, it is worth noting that in the case of the merger Abell 520 the dark matter coincides with the plasma.  If the dark matter was stopped by a self-interaction, the cross section for this process would need to be appreciably larger than the upper bound established using the bullet cluster \cite{bulletcattivo}.  Similarly the dark matter profile of a galaxy cluster depends  upon its unknown formation history.  We have chosen not to discuss type Ia supernovae as the standardizability of these candles is an empirical observation, still not supported by a calculation, and so again the errors cannot be estimated.  Indeed, assuming $\Lambda$CDM the cosmic distance ladder based on standardizable candles \cite{hst,hst2} is now in about 3$\sigma$ disagreement with CMB observations \cite{planck}.

However one cosmological probe is independent of any known systematics or cosmological assumptions, the baryonic acoustic oscillation peak (BAO) \cite{bao1}.  This is the location of an isolated peak in the position space two point function of the matter distribution.  As the matter moves nonrelativistically, the location of this peak in comoving coordinates is time independent.  Thus, while the size itself depends on the cosmological model, once the size is measured at any redshift, it will be the same in comoving coordinates at all redshifts.  Furthermore, the location of the peak is isotropic in the sense that the peak of the matter distribution in the radial and angular directions at the same redshift agree.  Thus at each redshift one can measure two quantities, the angular and radial BAO size.  The knowledge that each scale is the same in comoving coordinates then allows one to robustly determine the expansion of the universe as a function of time.  Using Einstein's equations this rate of expansion yields the pressure and density of the universe as a function of time.  If one further assumes that the universe consists of various components whose stress energy tensors are separately conserved and that dark matter corresponds to a nonrelativistic fluid $(w=0)$ then one obtains the average dark matter density at each time.   

The dark matter density so obtained, going back to about $z=2.3$ \cite{baolyman}, is consistent with that obtained from the CMB \cite{planck} corresponding to redshifts going back to $10^4$.  This implies that the quantity of dark matter in a sufficiently large comoving volume should have been roughly constant since redshift $z=10^4$.  While this result is inconsistent with modified gravity models of dark matter, it is consistent with models of stable particles or other such compact objects.

\section{Challenges faced by WIMPs} \label{challsez}

The successes of WIMPs have one thing in common.  They are all at large distance scales, comoving scales of Mpc.  What about short distance predictions?

One robust prediction of CDM WIMPs is that the Milky Way should have at least of order $10^4$ satellite dark matter halos \cite{klypin,moore}.  Only about 25 such satellites have so far been observed, in particular the lightest ones seem to be missing.  In itself this is not a contradiction, current lensing probes are not yet sufficiently sensitive to observe these halos.  It may be that, since the lightest halos lead to shallow gravitational potential wells, the primordial gas in these wells was blown away by ultraviolet radiation from reionization \cite{uv1,uv2} or cosmic ray pressure \cite{cr} before it could condense into stars, or that the supernovas of the first generation of stars blew out of all the gas so that no further generations formed \cite{snfeedback}.  Whether these mechanisms really would have suppressed star formation in these light halos depends on many unknowns, such as the ultraviolet and cosmic ray fluxes in the distant past and the efficiency with which supernovae can transfer their energy to the gas.  Even if these mechanisms may evade star formation in small satellites and so explain the nondetection of light halos, there also appear to be missing heavy halos \cite{toobig} for which such mechanisms are of no avail \cite{garrison}.   One can eliminate these heavy halos from simulations by reducing the mass of the Milky Way \cite{mwleggera,mwleggera2} but this leads to tension with theoretical models of the mass distribution of the Milky Way~\cite{mcmillan}.  Also such a low Milky Way mass would imply that the dwarf galaxy Leo I \cite{leounbound} and the Magellanic clouds \cite{magunbound} are gravitationally unbound, which makes their observed proximity to the Milky Way unlikely.

Pure dark matter simulations yield a second prediction, that the density profile of a dark matter halo diverges at small radii \cite{nfw}.  Such divergent halos are said to be cusped.  Recent simulations \cite{aquarius} suggest that the power law divergence be replaced by a more complicated Einasto form \cite{einasto}, but in practice the diminishing divergence occurs at such small radii that it will not have any effect on the discussion below.  Observed halos do not appear cusped, but they are not pure dark matter, they also contain stars and often gas.  Could the baryonic physics of stars eliminate the cusps?  A number of simulations have indicated that in principle the outflow of gas from a supernova may turn a cusped profile into a cored profile \cite{sn1,sn2}.  However this transformation only works in the supernovae is able to efficiently transfer its energy, which requires these models to use a very high threshold gas density for star formation \cite{troppogas}.   There has been considerable debate concerning whether such a threshold is reasonable.  

While it may be that baryonic physics can transform the cusps into cores in some dark matter halos, clearly such a transformation is impossible if the baryon density is insufficient.  The main proponents of this mechanism \cite{gov2012} find, as summarized in their Eq.~(1), that it can totally remove cusps in galaxies with more than $10^9\ M_\odot$ of stars whereas the central density diverges at least as $1/r$ if the mass is beneath $4\times 10^ 6\ M_\odot$.  Thus smaller galaxies, such as the majority of the dwarf spheroidal galaxies in our local group, will unavoidably have cusped dark matter halos if dark matter is made of WIMPs.  Are these cusps in contradiction with observations?  These light galaxies are not rotating, so their rotation curves can not be used to determine their dark matter halo profiles.   The stars are dispersion supported, so in principle the halo profiles can be determined from the Jeans equation \cite{jeans}.  However, as a result of a degeneracy \cite{jeansdegen} this is not sufficient to determine the halo density.  Instead, if a few basic assumptions are included \cite{wolf}, the Jeans equation yields the amount of mass within the half-light radius \cite{walker} of each stellar population \cite{pops}.  Applying this method to galaxies with multiple stellar populations one can obtain the densities at distinct radii.  This has been done for the Fornax \cite{pops} and Sculptor \cite{sculptor} dwarfs and in both cases the result is consistent with a cored profile and not a cusped profile, although the stellar mass of Sculptor is only  $2\times 10^ 6\ M_\odot$ and so the results of \cite{gov2012} suggest a cusped profile.  Furthermore, assuming it not to be recently accreted, the existence of old substructure in the Fornax \cite{fornax}, Ursa Minor \cite{ursaminor} and Sextans \cite{sextans} dwarfs is inconsistent with a cusped profile, as the resulting tidal force would both delocalize the substructure and also pull it towards the center of the halo.

Recently WIMP models have faced new challenges arising from the concentration in phase space of dwarf satellites of the Milky Way \cite{milkydisco} and Andromeda \cite{andromedadisco} as well as the planes of Ref.~\cite{duedischi} and the filamentary structure of \cite{filament} and the abundance of pairs in Refs.~\cite{fattahi,paii}.  In CDM WIMP simulations such concentrations do not appear, although perhaps they may be caused by recent mergers \cite{belreview}.

\section{Proposal: giant monopoles}

The standard model contains no dark matter candidates which are consistent with the tight bounds on MACHOs.  As a result, any explanation of dark matter must introduce either new and exotic couplings or else new fields.  We choose the later path, introducing a dark SU(2) gauge field, a scalar Higgs field which transforms in the adjoint representation of this dark SU(2) and also an undetermined number of fundamental fermions.  These fields are coupled via the same gauge-invariant and renormalizable terms that appear in the standard model.  These fields all inhabit a dark sector, they are in addition to the usual standard model fields.  They will interact with the standard model fields gravitationally, although renormalizability and gauge invariance also allow the Higgs portal interaction of Ref.~\cite{kiasmonopole}.

This dark sector admits a stable classical field theory solution, the 't Hooft-Polyakov monopole.  While it is known that a small fraction of dark matter consists of active, massive neutrinos, following Ref.~\cite{bjarke}, we will propose that the rest of the dark matter in our universe consists of these monopoles.  Each monopole, in a steady state and up to a discrete choice that we will discuss below, is completely characterized by a single integer, its charge $Q$.  We will propose that each dark matter halo consists of a {\it single} monopole of some charge $Q$, thus recovering the observational fact that galaxies tend to inhabit one parameter families.  This observation has consistently posed a challenge to WIMP models which naturally yield two parameters, corresponding intuitively to a mass and a temperature.  Note that, since halos are extremely large, so are the monopoles. The monopoles will have diameters measured in parsecs and masses measured in solar masses, in contrast with the more familiar and smaller GUT monopoles.  In the rest of this talk we investigate the consequences of this proposal.

\section{Giant monopoles at large scales}

Does the giant monopole model reproduce the successes of WIMPs at large scales?  At scales much larger than the monopole, the monopoles can be treated as point particles.  As we will discuss later, at long distances only their gravitational interactions will be significant.  Therefore their behavior will be indistinguishable from WIMPs.  In particular, the profiles of galaxy clusters will be the same as will the relative BAO scales.  

The overall BAO scale is determined by the speed of sound of the primordial plasma, which is unchanged in our scenario, and by the age of the universe at recombination.  What is the age of the universe at recombination in this model?  First of all we can calculate the time when the monopoles form using the usual Kibble mechanism.  The monopoles form when the scalar Higgs field unfreezes, which occurs when the Hubble length grows beyond $r_1$, the inverse tachyonic mass of the Higgs field.  The scale $r_1$ is just the radius of the core of a dark matter halo which is about 1 kpc, so monopoles form when the Hubble radius is about one inverse kpc.  This is before matter-radiation equality, so before the monopoles have formed the dark matter, in either WIMP or monopole models, is an insignificant part of the total energy budget of the universe.  At matter-radiation equality, when the contribution of dark matter to the energy density of the universe becomes relevant, the monopoles have already formed and so, as the monopoles are stable, it is reasonable that the mass in each comoving volume is then fixed as in a WIMP cosmology.  In this case the energy density of monopoles before recombination is the same as that of WIMPs in the standard cosmological model, and so recombination happens at the same moment.  This implies that the overall scale of the BAO feature will be the same as in WIMP cosmologies, in agreement with observations.  Note that, had the cores of galaxies today been larger by a factor of ten, the monopoles would have formed too late and this model would have been falsified.

What about the CMB?  As the monopoles have formed before matter-radiation equality, they may contribute the evolution of the power spectrum.  Once they are fully formed, they will have the correct nonrelativistic equation of state $w=0$, although at early times $w$ begins at $-1$.  Were they a fluid?  Evolving back the density of dark matter today one finds about $10^2$ monopoles in each $l=2000$ volume \cite{bjarke}, just enough for the fluid approximation at these multipoles.

%, which depends on the total dark matter mass between matter-radiation equality and recombination.  The total mass in a comoving volume is constant in the case of both monopoles and WIMPs, as the particles are conserved.  Therefore the overall BAO scale is identical to that in a WIMP cosmology so long as the monopoles have already formed by matter-radiation equality.  When do the monopoles form?  As in the usual Kibble mechanism, the monopoles form when the Hubble radius is of order the tachyonic mass of the Higgs field.  We will see below that this is equal to the inverse radius of the core of an individual monopole solution, which is of order one kpc.  Thus the monopoles form when the Hubble radius is of order 1 kpc, which is somewhat earlier than matter-radiation equality.

The monopoles are particles with inertia and so, as is observed in the bullet cluster, will tend to continue straight through a collision unless scattered.   The bullet cluster bound on the cross section \cite{bulletlim} of about 1 cm${}^2$/g, for a $10^{12} M_\odot$ halo, corresponds to the cross section of a black disk of radius equal to about 5 kpc.  While we have not yet computed the cross sections for our solutions, the dark force interactions will be stronger than the gravitational interactions by only about 5 orders of magnitude and so it is plausible that this bound is satisfied.

\section{Giant monopoles at small scales}
We have seen that it is plausible that giant monopoles share the same large scale successes as WIMPs, although this is contingent on details of their formation before recombination.  The real strength of the monopole models is at small scales, where they effortlessly solve the challenges described in Sec.~\ref{challsez}.  First of all, the missing satellite problem.  The missing light satellite problem is essentially the observation that satellite galaxies are observed to have a minimum mass, in contradiction with the universal expectation from WIMP simulations.  On the other hand, Dirac has shown that monopole solutions always have a minimum mass, corresponding to the $Q=1$ monopole.  Below we will fix the parameters of our theory such that the $Q=1$ monopole has the observed minimum mass, and thus the missing halo problem will be solved.

What about the cusp problem?  The profile of a 't Hooft-Polyakov monopole in the presence of gravity, at least in the case $Q=1$, is well known.  There is a central core, of radius $r_1$ with a reasonably constant density and a nearly zero Higgs and gauge field.  Thus the core problem is automatically solved.  At the radius $r_1$ the Higgs field turns on, as $r_1$ is proportional to the inverse Higgs mass, breaking the SU(2) gauge symmetry to U(1).  Let $r_2$ be the inverse W boson mass, where the W boson is not the standard model W boson but the gauge field which has been Higgsed.  Then, between $r_1$ and $r_2$ the gauge fields are essentially zero but the Higgs field enjoys a topologically nontrivial winding which gives this region a density proportional to $1/r^2$.  This automatically yields the observed flat rotation curves.  At higher radii the gauge fields also become nontrivial.  

And the phase space correlations?  The dark forces in these models are more attractive than gravity in some regimes, and so one expects more binding and so more correlations.  Whether this increase agrees with observations will be studied in the near future.

What are the parameters of the model?  Dwarf spheroidal galaxies (dSphs) are among the purest concentrations of dark matter in the universe.  By using Refs.~\cite{walker,wolf} to fix to their masses inside of their half-light radii and assigning each dSph a charge $Q$ we have fit the Higgs VEV to be $v\sim 10^{14}$ GeV, the quartic Higgs selfcoupling to be $\lambda\sim 10^{-96}$ and the gauge coupling to be less than $10^{-48}$.   The value of $v$ is tantalizing, the inputs were astrophysical scales and the output is the leptogenesis scale.

\section{Challenge}
Unfortunately these monopoles repel and so the charge $Q>1$ halos are unstable.  This may rule out our model.  Then again, protons repel but visible matter is mostly made of protons, as the repulsion at small distances is canceled by neutrons and at large distances is screened by electrons.  The monopoles only repel at long distances.  So what are the analogs of the electrons?  Electrons carry the opposite charge from protons but cannot annihilate with protons as they carry a flavor quantum number and the lightest state for a decay product, the neutron, is too massive for the decay to be kinematically allowed.  Similarly such a flavor quantum number for the monopoles is an automatic consequence of our fermionic couplings \cite{jr}.  The masses of the various flavors of monopoles can be adjusted by choosing the Yukawa couplings.  We propose to include light antimonopoles of a different flavor which screen the long distance repulsion of our monopoles.  If such a screening cannot be made to work, our proposal will be excluded.

\ack
Many of the results presented in this talk were obtained in collaboration with Sven Bjarke Gudnason.  JE is supported by the Chinese Academy of Sciences Fellowship for
Young International Scientists grant number 2010Y2JA01.

\end{document}